\documentclass{PoS}

\usepackage{amsmath,amstext,amsfonts,amsbsy,amssymb,amscd,bbm,epsfig,lscape}
\newcommand{\ba}{\begin{array}}
\newcommand{\ea}{\end{array}}

\newcommand{\refig}[1]{Fig.~\ref{#1}}

\newcommand{\dif}{{\rm d}}

\newcommand{\Dslash}{\relax{\kern+.25em / \kern-.70em D}}

\newcommand{\MeV}{{\rm MeV}}

\newcommand{\Real}{\relax{\mathsf{\Gamma\kern-.35em R}}}
\newcommand{\Int}{\relax{\mathsf{Z\kern-.40em Z}}}

\newcommand{\NF}{N_\mathrm{\scriptstyle f}}

\newcommand{\gbar}{\kern1pt\overline{\kern-1pt g\kern-0pt}\kern1pt}
\newcommand{\mbar}{\kern2pt\overline{\kern-1pt m\kern-1pt}\kern1pt}
\newcommand{\obar}[1]{\kern3pt\overline{\kern-2pt #1\kern-0pt}\kern1pt}

\newcommand{\lmax}{L_{\rm max}}

\newcommand{\Oa}{\mbox{O}(a)}

\newcommand{\abar}{\kern1pt\overline{\kern-1pt a\kern-0.5pt}\kern1pt}

\newcommand{\cO}{{\cal O}}

\newcommand{\cQ}{{\cal Q}}

\title{Non-perturbative renormalisation of four-fermion operators in $\NF=2$ QCD\thanks{CERN-PH-TH/2007-178, FTUAM-07-14, IFT-UAM-CSIC-07-45.}}

\ShortTitle{NPR of four-fermion operators in $\NF=2$ QCD}

\author{\epsfig{figure=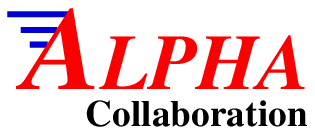,width=25truemm}}

\author{Petros Dimopoulos, Gregorio Herdoiza\thanks{Address after Oct 1st: DESY-Zeuthen, Platanenallee 6, D-15738 Zeuthen, Germany.}, Anastassios Vladikas\\
        INFN, Sezione di Roma ``Tor Vergata'' and Dipartimento di Fisica, Universit\`a di Roma ``Tor Vergata''\\
        Via della Ricerca Scientifica 1, I-00133 Rome, Italy\\
        email: \email{petros.dimopoulos@roma2.infn.it, gregorio.herdoiza@roma2.infn.it, tassos.vladikas@roma2.infn.it}}

\author{Filippo Palombi, Mauro Papinutto\\
        CERN, Physics Department, Theory Division\\
        CH-1211 Geneva 23, Switzerland\\
        email: \email{filippo.palombi@cern.ch, mauro.papinutto@cern.ch}}

\author{\speaker{Carlos Pena}\\
        Departamento de F\'{\i}sica Te\'orica and
        Instituto de F\'{\i}sica Te\'orica UAM/CSIC\\
        	Facultad de Ciencias, Universidad Aut\'onoma de Madrid,
		Cantoblanco E-28049 Madrid, Spain\\
        email: \email{carlos.pena@uam.es}}

\author{Hartmut Wittig\\
        Institut f\"ur Kernphysik, University of Mainz\\
        D-55099 Mainz, Germany\\
        email: \email{wittig@kph.uni-mainz.de}}

\abstract{We present results for the non-perturbative renormalisation of four-fermion operators with two flavours of dynamical quarks. We consider both fully relativistic left current-left current operators, and a full basis for $\Delta B=2$ operators with static heavy quarks. The renormalisation group running of the operators to high energy scales is computed in the continuum limit for a family of Schr\"odinger Functional renormalisation schemes, via standard finite size scaling techniques. The total renormalisation factors relating renormalisation group invariant to bare operators are computed for a choice of lattice regularisations.}

\FullConference{The XXV International Symposium on Lattice Field Theory\\
		 July 30-4 August 2007\\
		 Regensburg, Germany}

\begin{document}

\section{Introduction}

Hadronic matrix elements (HMEs) of four-fermion operators have long been essential input quantities for Flavour Physics. Reliable estimates of a number of HMEs
are crucial in the study of CP violation via CKM unitarity triangle analyses,
or of such striking experimental findings as the enhancement of hadronic
decay amplitudes by long-distance effects (as e.g. in the $\Delta I=1/2$ 
rule). The only known technique to compute HMEs from first
principles is lattice QCD. However, lattice QCD results have long
been hampered by the difficulty to eliminate a number of systematic 
uncertainties. Most notably, the high cost of including  dynamical quark 
effects in lattice QCD simulations has enforced for many years either the 
quenched  approximation, or the use of dynamical quark masses far too heavy to 
allow for a well-controlled extrapolation to the physical regime. In some
cases, e.g. the computation of the kaon bag parameter $B_K$, quenching effects are indeed the last remaining uncontrolled systematic uncertainty~\cite{review}.

As techniques for the simulation of light dynamical quarks have witnessed
dramatic progress in the last few years (see e.g.~\cite{Giusti:2007hk}), it becomes increasingly important
to bring to this environment the techniques to control other sources of
uncertainty, in order to aim at precision computations of physical quantities.
In the context of HMEs, one of the most prominent examples is non-perturbative
renormalisation (NPR) (see e.g.~\cite{Sommer:2002en}). The use of finite-size scaling techniques has
allowed to control fully both the renormalisation group (RG) running and the
matching of lattice to renormalised observables in the quenched approximation
for a broad class of four-fermion operators~\cite{Guagnelli:2005zc,Palombi:2005zd,Palombi:2007dr}. The aim of
the present work is to extend these results to $\NF=2$ QCD. In particular,
we will discuss 1. the RG running of left current-left current relativistic
four-fermion operators, 2. the RG running of all $\Delta B=2$ operators with static heavy quarks, and 3. the matching of the above operators to
renormalised continuum 
operators for some particular choices of the regularisation. Immediate
applications, as we will point out later, arise in the
computation of the bag parameters $B_K$ and $B_B$. Preliminary results
had been presented at last year's conference~\cite{Dimopoulos:2006es}.

\section{Definitions and setup}

\subsection{Renormalisation of four-fermion operators}

We will consider two different classes of operators:
\begin{align}
O^\pm_{\Gamma_1,\Gamma_2}(x) &= \frac{1}{2}\left[
 \left(\bar\psi_1(x)\Gamma_1\psi_2(x)\right)
 \left(\bar\psi_3(x)\Gamma_2\psi_4(x)\right)
 \pm ( 2\leftrightarrow 4) \right] \,, \\
\cO^\pm_{\Gamma_1,\Gamma_2}(x) &= \frac{1}{2}\left[
 \left(\bar\psi_h(x)\Gamma_1\psi_2(x)\right)
 \left(\bar\psi_{\bar h}(x)\Gamma_2\psi_4(x)\right)
 \pm ( 2\leftrightarrow 4) \right] \,.
\end{align}
In the above expressions $\psi_k$ is a relativistic quark field with
flavour index $k$, $\psi_{h,\bar h}$ are static (anti)quark fields,
$\Gamma_l$ are spin matrices, and the parentheses indicate spin-colour traces.
All the fields are interpreted to be in the valence sector of the theory.
This formalism of distinct quark flavours will allow us to isolate
scale-dependent logarithmic divergences from eventual mixing with
lower-dimensional operators that may appear for specific choices of
quark masses and/or flavour content.

The above operators mix under renormalisation as determined by the symmetries
of the regularised theory. If we restrict ourselves to the parity-odd sector,
complete bases of operators in the relativistic and static cases are
given by
\begin{gather}
Q^\pm_k \in \left\{
O^\pm_{\rm\scriptscriptstyle VA+AV},
O^\pm_{\rm\scriptscriptstyle VA-AV},
O^\pm_{\rm\scriptscriptstyle SP-PS},
O^\pm_{\rm\scriptscriptstyle SP+PS},
O^\pm_{\rm\scriptscriptstyle T \tilde T},
\right\} \,;~~~~~
\cQ^\pm_k \in \left\{
\cO^\pm_{\rm\scriptscriptstyle VA+AV},
\cO^\pm_{\rm\scriptscriptstyle VA-AV},
\cO^\pm_{\rm\scriptscriptstyle SP-PS},
\cO^\pm_{\rm\scriptscriptstyle SP+PS},
\right\} \,,
\end{gather}
respectively,
in standard self-explanatory notation for the choice of spin matrices $\Gamma_l$. A full analysis of the renormalisation of these
operator bases with relativistic Wilson fermions has been performed in~\cite{Donini:1999sf,Palombi:2006pu}. One
particular conclusion is that, contrary to the parity-even case, discrete symmetries
protect all the above operators from extra mixings under renormalisation
due to the breaking of chiral symmetry. Recall that the RG of these
operators and of their parity-even partners is identical, as in the
continuum limit (CL) chiral symmetry holds. On the other hand, the connection to observables
involving matrix elements of parity-even operators is non-trivial.

From now on, we will consider the subset of operators
\begin{gather}
Q_1^\pm \,,~~
\cQ^{'+}_k \in \left\{ 
\cQ^+_1,\cQ^+_1+4\cQ^+_2,\cQ^+_3+2\cQ^+_4,\cQ^+_3-2\cQ^+_4
\right\} \,.
\end{gather}
All these operators renormalise multiplicatively --- i.e., given
an operator $O\in\{Q_1^\pm,\cQ^{'+}_k\}$ the corresponding operator
insertion in any on-shell renormalised correlation function is given by
\begin{gather}
O_{\rm R}(x;\mu) = \lim_{a \to 0} Z(g_0,a\mu)\,O(x;g_0) \,,
\end{gather}
where $g_0,a$ are the bare lattice coupling and the lattice spacing,
respectively. The RG running of the operator is controlled by the
anomalous dimension $\gamma$, defined by the Callan-Symanzik equation
\begin{gather}
\label{eq:CS}
\mu\frac{\partial}{\partial\mu}O_{\rm R}(x;\mu) =
\gamma(\gbar(\mu))\,O_{\rm R}(x;\mu) \,,
\end{gather}
which is supplemented by the corresponding Callan-Symanzik equation
for the renormalised coupling
\begin{gather}
\mu\frac{\partial}{\partial\mu}\gbar(\mu) = \beta(\gbar(\mu))\,.
\end{gather}
In mass-independent renormalisation schemes, the beta function and all
anomalous dimensions do indeed depend only on the renormalised coupling 
$\gbar$. They admit perturbative expansions of the form
\begin{gather}
\beta(g) \stackrel{g\to 0}{\approx} -g^3\left(b_0+b_1g^2+b_2g^4+\ldots\right) \,;~~~~~~~
\gamma(g) \stackrel{g\to 0}{\approx} -g^2\left(
\gamma_0 + \gamma_1g^2 + \gamma_2g^4 + \ldots\right) \,,
\end{gather}
in which the coefficients $b_0,b_1,\gamma_0$ are renormalisation scheme-independent. Upon formal integration of Eq.~(\ref{eq:CS}), one is left
with the renormalisation group invariant (RGI) operator insertion
\begin{gather}
\label{eq:rgi}
\hat O(x) = O_{\rm R}(x;\mu)\,
\left[\frac{\gbar^2(\mu)}{4\pi}\right]^{-\frac{\gamma_0}{2b_0}}
\exp\left\{
-\int_0^{\gbar(\mu)}\dif g\,\left(\frac{\gamma(g)}{\beta(g)}-
\frac{\gamma_0}{b_0g}\right)
\right\} \,,
\end{gather}
while the RG evolution between two scales $\mu_1,\mu_2$ is given
by the operator
\begin{gather}
\label{eq:evol}
U(\mu_2,\mu_1) = \exp\left\{
\int_{\gbar(\mu_1)}^{\gbar(\mu_2)}
\dif g \, \frac{\gamma(g)}{\beta(g)}
\right\} =
\lim_{a \to 0} \,\frac{Z(g_0,a\mu_2)}{Z(g_0,a\mu_1)} \,.
\end{gather}

\subsection{Schr\"odinger Functional renormalisation schemes}

Eq.~(\ref{eq:evol}) is the starting point to compute non-perturbatively
the RG evolution of composite operators. To that purpose we introduce
a family of Schr\"odinger Functional (SF) renormalisation schemes. The
latter are defined by regularising the theory on a symmetric lattice of 
physical size $L^4$ with SF boundary conditions (see e.g.~\cite{Sommer:2006sj} for an introduction to the SF setup).
The renormalisation scale is set to be the infrared cutoff,
i.e. $\mu=1/L$. Renormalisation conditions for relativistic operators have the 
form
\begin{gather}
\label{eq:rencon}
Z(g_0,a\mu) \,\frac{F(x_0)}{\Theta} =
\left.\frac{F(x_0)}{\Theta}\right|_{\rm tree~level} \,,
\end{gather}
and are imposed in the chiral limit.
In the above expression, $F$ is a four-point correlation function of the form
\begin{align}
F(x_0) = \frac{1}{L^3}\langle
\cO_{21}[\Gamma_A]\cO_{45}[\Gamma_B]Q^\pm_k(x)\cO'_{53}[\Gamma_C]
\rangle \,,
\end{align}
where $\cO[\Gamma]$ are bilinear interpolating fields living on the time
boundaries, and $\Theta$ is a suitable boundary-to-boundary correlation
function that divides out the ultraviolet divergences associated to
these bilinears. Similar renormalisation conditions to Eq.~(\ref{eq:rencon})
are set up for static-light operators, with flavours $1$ and $3$ substituted
by $h$ and $\bar h$. Full details are provided in~\cite{Guagnelli:2005zc,Palombi:2005zd,Palombi:2006pu}. For now
it is just important to mention that the renormalisation scheme is fully determined by
fixing the parameters involved in the SF boundary conditions;
the point $x_0$ at which Eq.~(\ref{eq:rencon}) is imposed;
the Dirac matrices $\Gamma_{A,B,C}$ entering boundary bilinears~\footnote{At vanishing external momenta, there are 5 possible nontrivial choices that
preserve cubic symmetry.}; and
the normalisation factor $\Theta$.
Specific schemes have been introduced in~\cite{Guagnelli:2005zc,Palombi:2005zd,Palombi:2006pu}. Here we will concentrate
in the cases which have been found to be best behaved in the quenched study,
namely scheme~$1$ for $Q^+_1$ and scheme~$8$ for $Q^-_1$ in the notation of~\cite{Guagnelli:2005zc,Palombi:2005zd}, and
the reference schemes for static-light operators defined in~\cite{Palombi:2007dr}.

A crucial observation is that all the above renormalisation schemes are
mass-independent by construction, and the resulting renormalisation factors
are flavour-blind. It then follows that they can be used to remove the 
logarithmic divergences from any four-fermion operator with the considered structure, irrespective of
its specific flavour content, once eventual subtractions due to mixing
with equal- or lower-dimension operators have been properly performed.

\subsection{Step-scaling functions}

The basic objects to study the RG evolution of composite operators
non-perturbatively are the step-scaling functions (SSFs)
\begin{gather}
\Sigma(u,a/L) =
\left.\frac{Z(g_0,a/(2L))}{Z(g_0,a/L)}\right|_{\gbar^2(1/L)=u} \,,
\end{gather}
which can be computed at several values of the lattice spacing for fixed
physical size (inverse renormalisation scale) $L$. The corresponding values
of $\beta$ are indeed fixed by requiring that the renormalised SF coupling
$\gbar^2$, and hence $L$, are kept constant. It is then possible
CL extrapolation
\begin{gather}
\sigma(u) \equiv \lim_{a \to 0} \Sigma(u,a/L) =
\left.U((2L)^{-1},L^{-1})\right|_{\gbar^2(1/L)=u}\,.
\end{gather}

Once $\sigma(u)$ is known for several different values of the
squared gauge coupling $u$, it is possible to reconstruct the RG evolution
factor $U(\mu_{\rm had},\mu_{\rm pt})$ between two extreme scales
$\mu_{\rm had}$, in the range of a few hundred
$\MeV$, and $\mu_{\rm pt}$ in the high-energy regime. This in turn
allows to compute the RGI operator of Eq.~(\ref{eq:rgi}) in a way
free from large uncontrolled systematic uncertainties. It is enough
to consider the exponential on the rhs of Eq.~(\ref{eq:rgi}) evaluated at
$\mu=\mu_{\rm had}$, and split it as
\begin{gather}
\exp\left\{-\int_0^{\gbar(\mu_{\rm had})}\dif g\,\left(\frac{\gamma(g)}{\beta(g)}-
\frac{\gamma_0}{b_0g}\right)\right\} =
\exp\left\{-\int_0^{\gbar(\mu_{\rm pt})}\dif g\,\left(\frac{\gamma(g)}{\beta(g)}-
\frac{\gamma_0}{b_0g}\right)\right\} U(\mu_{\rm pt},\mu_{\rm had}) \,.
\end{gather}
The second factor on the rhs is known non-perturbatively, while the
first factor can be safely computed at NLO in perturbation theory, provided
the scale $\mu_{\rm pt}$ is high enough so as to render NNLO effects
negligible.

\section{Non-perturbative computation of the RG running}

SSFs have been computed using the non-perturbatively $\Oa$ improved
Wilson action, and a HYP2 action for static quarks, at six different values of the SF coupling, corresponding
to six different physical lattice lengths $L$. For each volume we have 
simulated at three  different values of the lattice spacing, corresponding
to lattices with $L/a=6,8,12$ (respectively $L/a=12,16,24$) for the
computation of $Z(L)$ (resp. $Z(2L)$). We used the $\NF=2$ configurations 
generated by the ALPHA Collaboration for the determination of the
RG running of the quark mass~\cite{DellaMorte:2005kg}. All the technical details concerning
the dynamical simulations are discussed in the mentioned work.

\begin{figure}[!t]
\vspace{55truemm}
\includegraphics{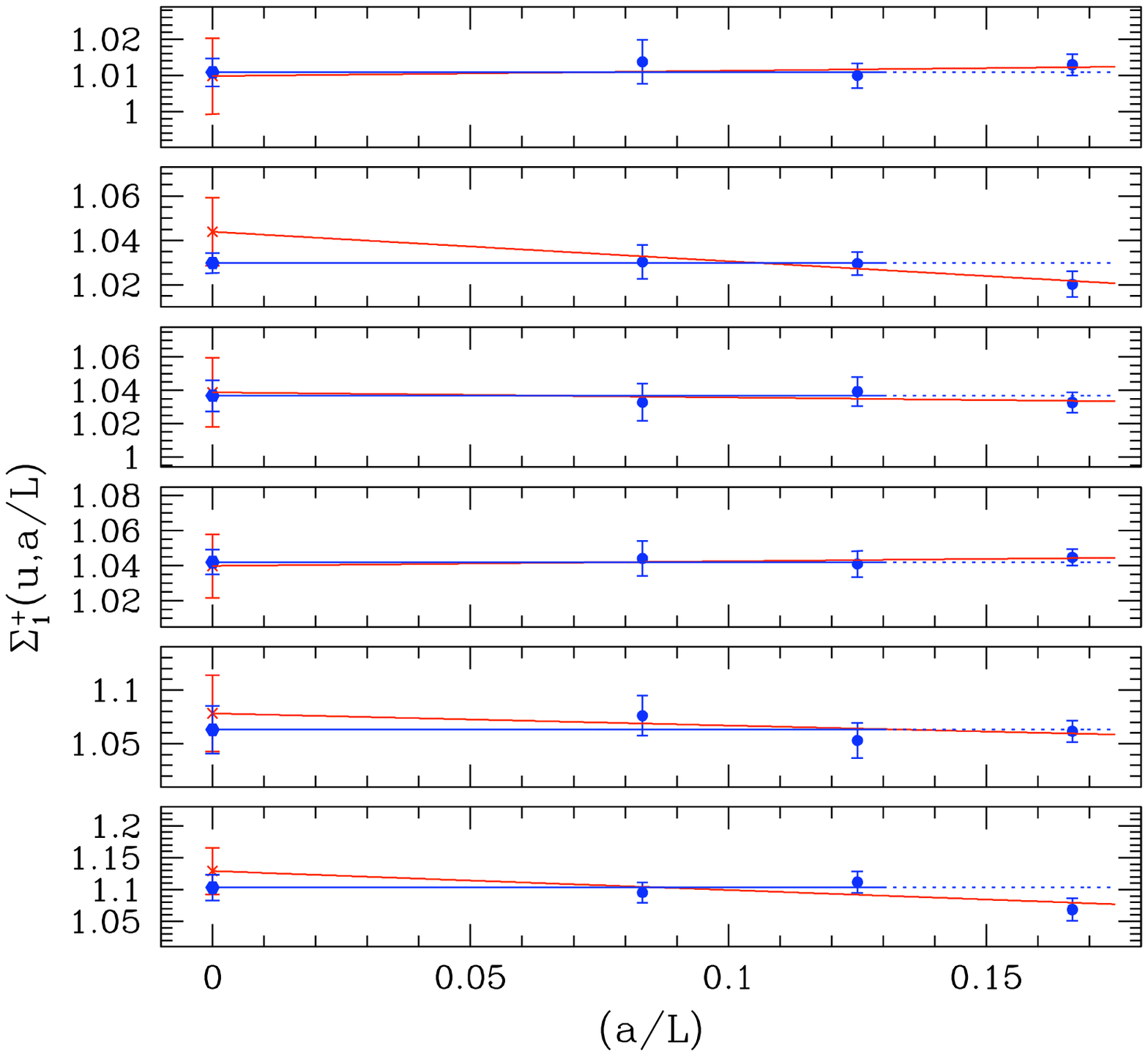}
\includegraphics{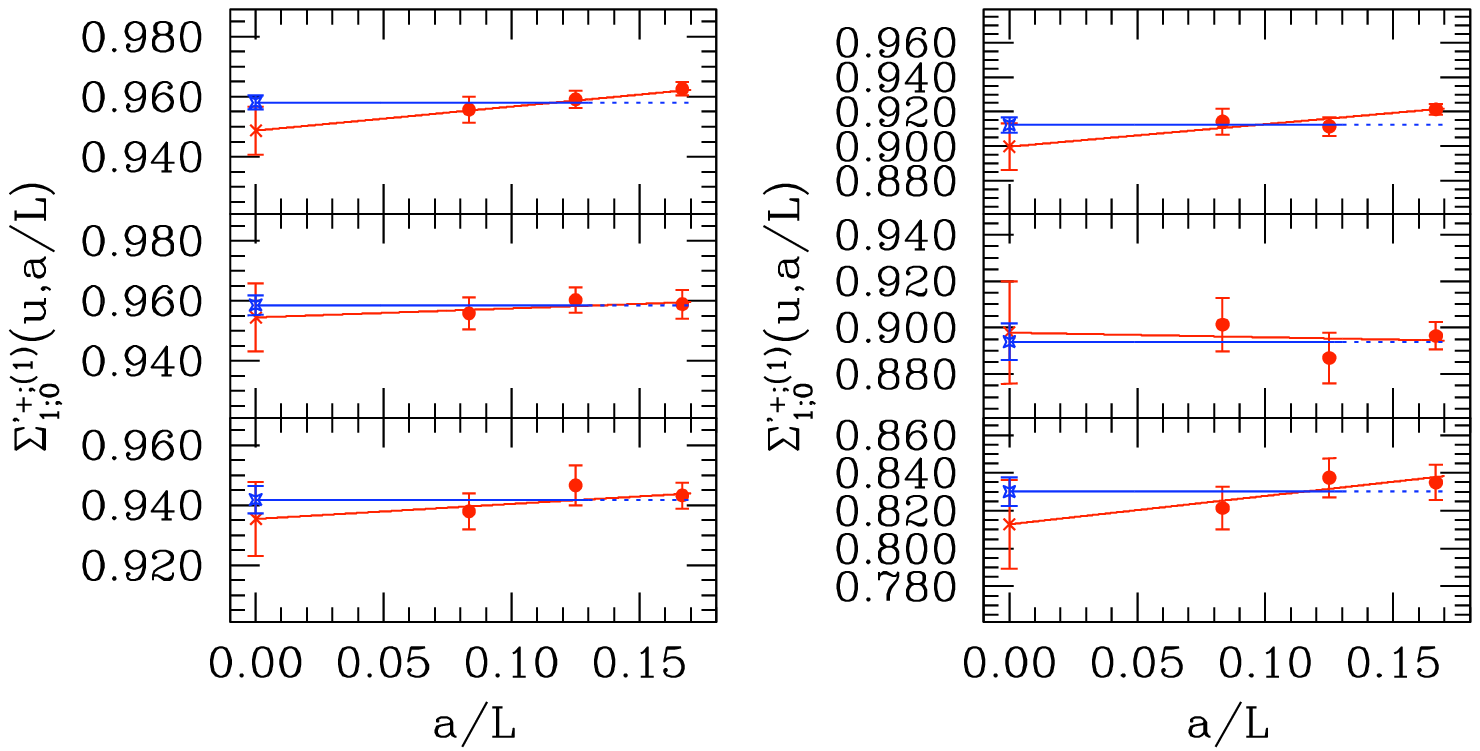}
\vspace{-0truemm}
\caption{CL extrapolation of the SSFs for $Q_1^+$ (left) and $\cQ_1^{'+}$ (right) for one particular choice of the renormalisation scheme (all boundary matrices set to $\gamma_5$, spatial boundary conditions set by
$\theta=0.5$, $\alpha=0$ in the renormalisation condition for $\cQ_1^{'+}$
(see~\cite{Guagnelli:2005zc,Palombi:2005zd,Palombi:2006pu} for details). The renormalised coupling increases
from top to bottom and from left to right. Blue discontinued lines and the blue point at
$a/L=0$ correspond to weighted averages of the $L/a=8,12$ data, red lines and the red $a/L=0$ cross to linear extrapolations in $a/L$ of the three data.}
\label{fig:extrap}
\end{figure}

As we do not implement full $\Oa$ improvement for four-fermion operators,
the only linear cutoff effects that are removed from $\Sigma(g_0,a/L)$
are those cancelled by the SW term in the fermion
action. Therefore, we expect SSFs to approach the CL linearly in $a/L$.
In practice, it is often observed that the data corresponding
to $L/a=8,12$ are compatible within errors, whereas the $L/a=6$ datum,
that is expected to bear the largest cutoff effect, is off.
This suggests that a weighted average of the results for the two finest
lattices, as considered in~\cite{DellaMorte:2005kg}, may yield a good estimate of the
CL value.
However, the lack of at least one extra value of $a/L$ closer to the continuum, that would 
allow a more precise control of the systematics, has led us to conservatively adopt linear
CL extrapolations involving all the data. It is worth remarking, though,
that linear fits and weighted averages lead to compatible results within one
standard deviation in most cases, as can be seen in \refig{fig:extrap}. The latter
illustrates the extrapolations at all values of the coupling
for two selected operators. Finally, let us mention that autocorrelation
times, which are included in the error estimate, increase towards the CL, leading to amplified errors in the finest lattices.

\begin{figure}[!t]
\vspace{42truemm}
\includegraphics{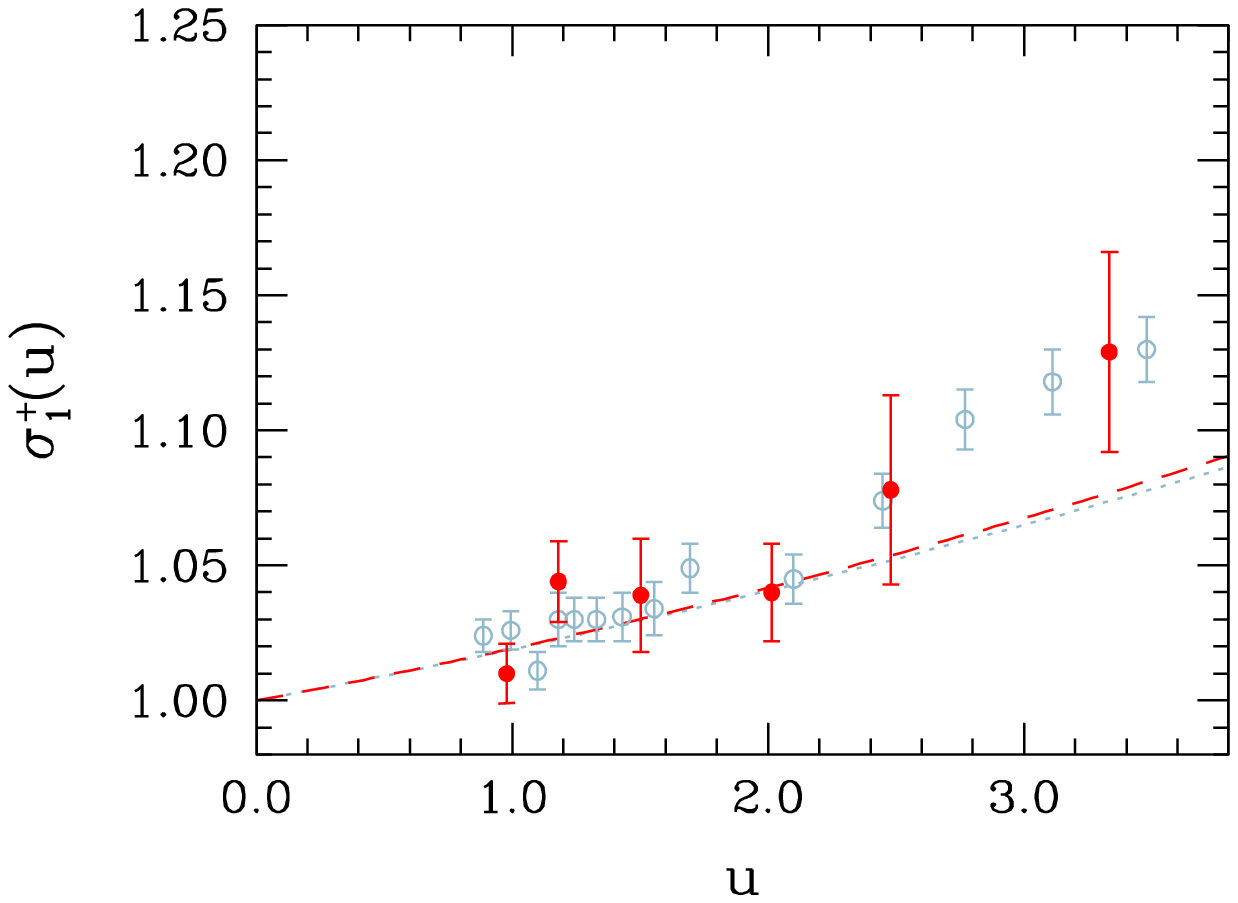}
\includegraphics{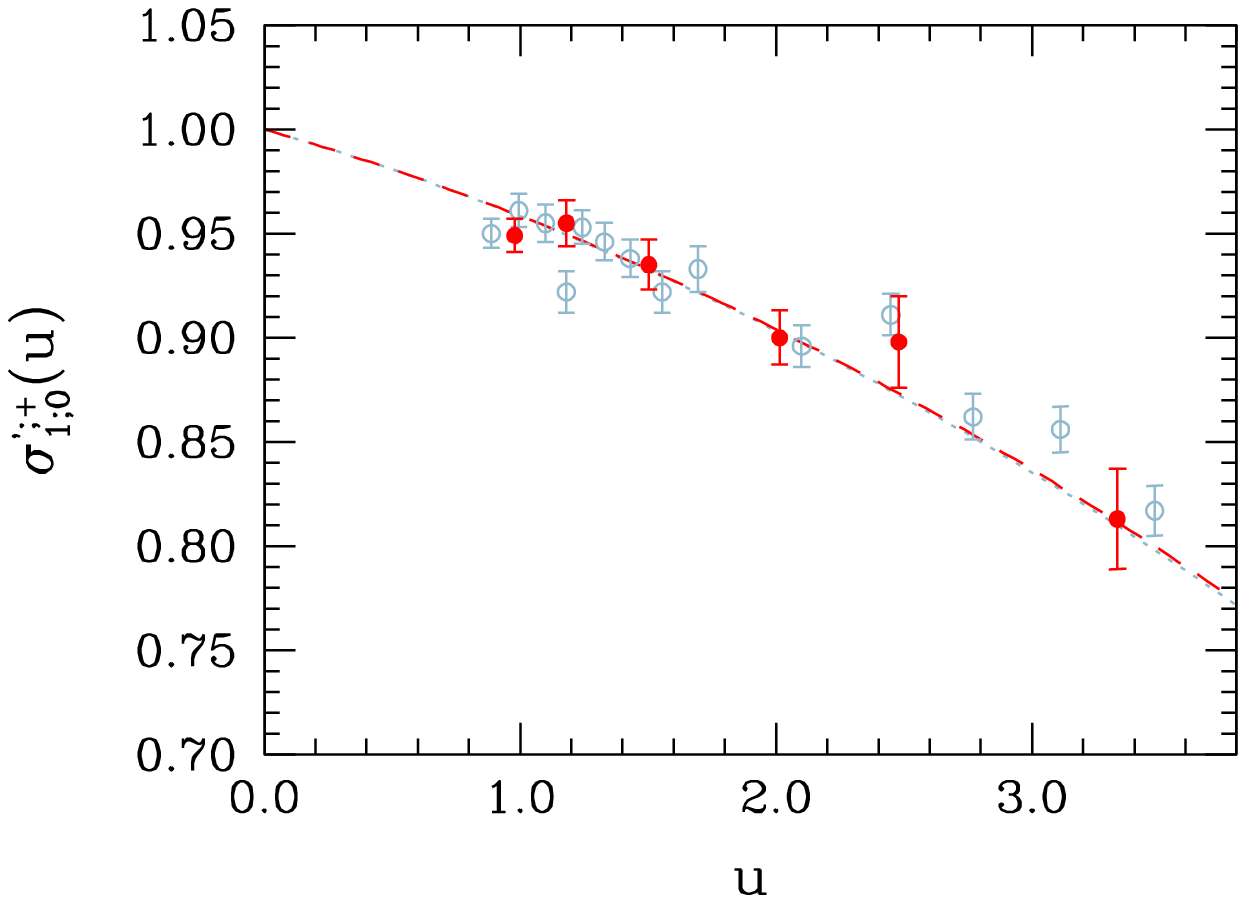}
\vspace{-0truemm}
\caption{SSFs for $Q_1^+$ (left) and $\cQ_1^{'+}$ (right) in the CL in the same
schemes as considered in Figure~1.
Full red points are $\NF=2$
results, open blue points are quenched results. The red dashed
(blue dotted) line is the NLO $\NF=2$ ($\NF=0$) perturbative result.}
\label{fig:CLssf}
\end{figure}

The resulting SSFs $\sigma(u)$ have been fitted to a polynomial form.
For definiteness, we will provide results for a fit to
$\sigma(u)=1+s_1u+s_2u^2+s_3u^3$, where $s_1$ is fixed at the value predicted
by $LO$ perturbation theory and $s_2,s_3$ are left as free parameters.
Once this continuous form of the SSF has been obtained, it is possible
to compute the relation between the RGI operators and the renormalised
operators at the low-energy scale $\mu_{\rm had}=\lmax^{-1}$, defined by
$\gbar^2(\lmax)=4.61$, as explained e.g. in \cite{Guagnelli:2005zc,Palombi:2007dr}.
This scale is chosen such that the renormalisation constant
$Z(g_0,a\mu_{\rm had})$ can be computed on accessible lattices
in ranges of values of $g_0$ commonly used in large volume simulations.
The results for the operators under investigation are reported in
Table~\ref{tab:ratios}. Note that typical relative errors reach
the 5\% ballpark, which may result in a sizeable error in HMEs
coming from renormalisation alone.

\section{Connection to hadronic observables}

RGI operator insertions can be related to bare operator insertions via
a total renormalisation factor $\hat Z(g_0)$, defined as
\begin{gather}
\hat Z(g_0) = Z(g_0,a\mu_{\rm had})\exp\left\{\int_0^{\gbar(\mu_{\rm had})}\dif g\,\left(\frac{\gamma(g)}{\beta(g)}-
\frac{\gamma_0}{b_0g}\right)\right\} \,.
\end{gather}
This is enough to remove all ultraviolet divergences, once eventual
renormalisation scale-independent mixing
with operators of dimension $d\le 6$ peculiar to the specific flavour
structure under consideration has been taken into account via suitable
subtractions. The details of the mixing depend on the regularisation
in which bare correlation functions are computed, as does the relation
between the latter and physical observables. For instance, in~\cite{tmbk,tmbb}
it has been explained how to extract the bag parameters $B_K$ and $B_B$
(the latter in the static limit for the $b$ quark) directly from three-point
functions involving the operators $Q_1^+$ and $\cQ_{1,2}^+$, by using
Wilson actions with suitable twisted mass terms. The computation of
the RGI renormalisation factors $\hat Z(g_0)$ at a number of values
of the bare coupling with an $\Oa$ improved Wilson action is under way
and close to completion.

\begin{table}[!t]
\vspace{-5mm}
\begin{center}
\begin{tabular}{cc@{\hspace{20truemm}}cc}
\hline\\[-2.5ex]
operator & ratio & operator & ratio \\[1.0ex]
\hline\\[-2.0ex]
         &             & $\cQ_1^{';+}$ & $0.724(34)$ \\
 $Q_1^+$ & $1.201(66)$ & $\cQ_2^{';+}$ & $0.647(32)$ \\
 $Q_1^-$ & $0.554(21)$ & $\cQ_3^{';+}$ & $0.539(18)$ \\
         &             & $\cQ_4^{';+}$ & $0.796(20)$ \\[0.5ex]
\hline
\end{tabular}
\end{center}
\caption{Ratios of RGI to renormalised at $\mu_{\rm had}$ operator insertions for the various operators in the reference renormalisation schemes mentioned in the text.}
\label{tab:ratios}
\end{table}

\section{Conclusions}

We have presented a fully non-perturbative computation of the RG running
of a wide class of four-fermion operators in $\NF=2$ QCD. These results,
together with the matching to specific hadronic schemes, is a basic
building block of any $\NF=2$ computation of such quantities as $B_K$
and $B_B$ that aims at eliminating systematic uncertainties related to
renormalisation. On the other hand, the precision of the results sets
a potentially unsatisfactory lower bound for the final error on weak
matrix elements. Future refinement, e.g. by adding a finest lattice
to our continuum limit extrapolations, can be hence desirable.
These issues will be discussed in detail in our forthcoming publication
of the definitive results.

\end{document}